\documentclass[jap,amsmath,amssymb,numerical,reprint,nofootinbib,letterpaper,superscriptaddress]{revtex4-2}
\usepackage{graphicx}
\usepackage{tabularx}
\usepackage{bm}
\usepackage{amsfonts,amsmath}
\usepackage{mathtools}
\usepackage{array}
\usepackage{xcolor}
\definecolor{steelblue}{rgb}{0.1,0.2,0.9}

\begin{document}
\setcitestyle{square}
\title{Superparamagnetic and Stochastic-Write Magnetic Tunnel Junctions for High-Speed True Random Number Generation in Advanced Computing}

\author{Jonathan Z. Sun}
\author{Christopher Safranski}
\author{Siyuranga Koswata}
\author{Pouya Hashemi}
\affiliation{IBM T. J. Watson Research Center,Yorktown Heights, NY 10598, USA}
\author{Andrew D. Kent}
\affiliation{Center for Quantum Phenomena, Department of Physics, New York University, New York, NY 10003, USA}

\date{\today}

\begin{abstract}
We review two magnetic tunnel junction (MTJ) approaches for compact, low-power, CMOS-integrated true random number generation (TRNG). The first employs passive-read, easy-plane superparamagnetic MTJs (sMTJs) that generate thermal-fluctuation-driven bit streams at 0.5--1~Gb/s per device. The second uses MTJs with magnetically stable free layers, operated with stochastic write pulses to achieve switching probabilities of about 0.5 (\emph{i.e.}, write error rates of $\simeq 0.5$), achieving $\gtrsim 0.1$~Gb/s per device; we refer to these as stochastic-write MTJs (SW-MTJs). Randomness from both approaches has been validated using the NIST~SP800 test suites. sMTJ approach uses a read-only cell with low power and can be compatible with most advanced CMOS nodes, while SW-MTJs leverage standard CMOS MTJ process flows, enabling co-integration with embedded spin-transfer torque magnetic random access memory (STT-MRAM). Both approaches can achieve deep sub-0.01~$\mu$m$^2$ MTJ footprints and offer orders-of-magnitude better energy efficiency than CPU/GPU-based generators, enabling placement near logic for high-throughput random bit-streams for probabilistic computing, statistical modeling, and cryptography. In terms of performance, sMTJs generally suit applications requiring very high data-rate random bits near logic processors, such as probabilistic computing or large-scale statistical modeling. Whereas SW-MTJs are attractive option for edge-oriented microcontrollers, providing entropy sources for computing or cryptographic enhancement. We highlight the strengths, limitations, and integration challenges of each approach, emphasizing the need to reduce device-to-device variability in sMTJs—particularly by mitigating magnetostriction-induced in-plane anisotropy—and to improve temporal stability in SW-MTJs for robust, large-scale deployment.
\end{abstract}

\maketitle

\flushbottom

\section{Introduction}

A nanoscale thin-film magnetic tunnel junction (MTJ) is the central element in spin-transfer-torque magnetic random access memory (STT-MRAM) offered by advanced CMOS foundries~\cite{IEDM23Forum,TMRC2024Forbes,2021150B,2021195,2017019}. In a similar technology setting, but with modified device structures, circuit designs, and operating modes, the same MTJ can serve as a compact, low-power, back-end-of-line (BEOL) CMOS-integrated source of entropy for advanced computing~\cite{2023010,2018146,2018149,2024126,2021104,2021140B,2022090,2022076,2023054,2024001,2024002}. In this review, we examine the key attributes, performance metrics, and integration challenges of MTJ-based entropy sources for advanced computing applications.

STT-MRAM is primarily used as an embedded memory in which perpendicularly magnetized MTJs (pMTJs)~\cite{2011014,2010058} are closely integrated directly above logic circuits, enabling data rates well matched to logic operations~\cite{2021005,2024128}. Using similar structures, stochastic MTJs can serve as fast (gigabit per second per MTJ -- Gbps/MTJ, or higher), high-density ($\ll 100$~nm MTJ diameter), and low-power random bit generators that are BEOL integrated with logic for workloads requiring large volumes of random data. Such high-quality random bit streams may also help hardware implementations of quantum-safe cryptography.

For these applications, the viability of an MTJ-based approach depends on several key attributes: (1) the speed of random bit-stream generation, (2) the statistical quality of the output sequence, and (3) the consistency and ease with which the bit-stream properties can be controlled.

\begin{figure}[tbp!]
\includegraphics[width=3.4in]{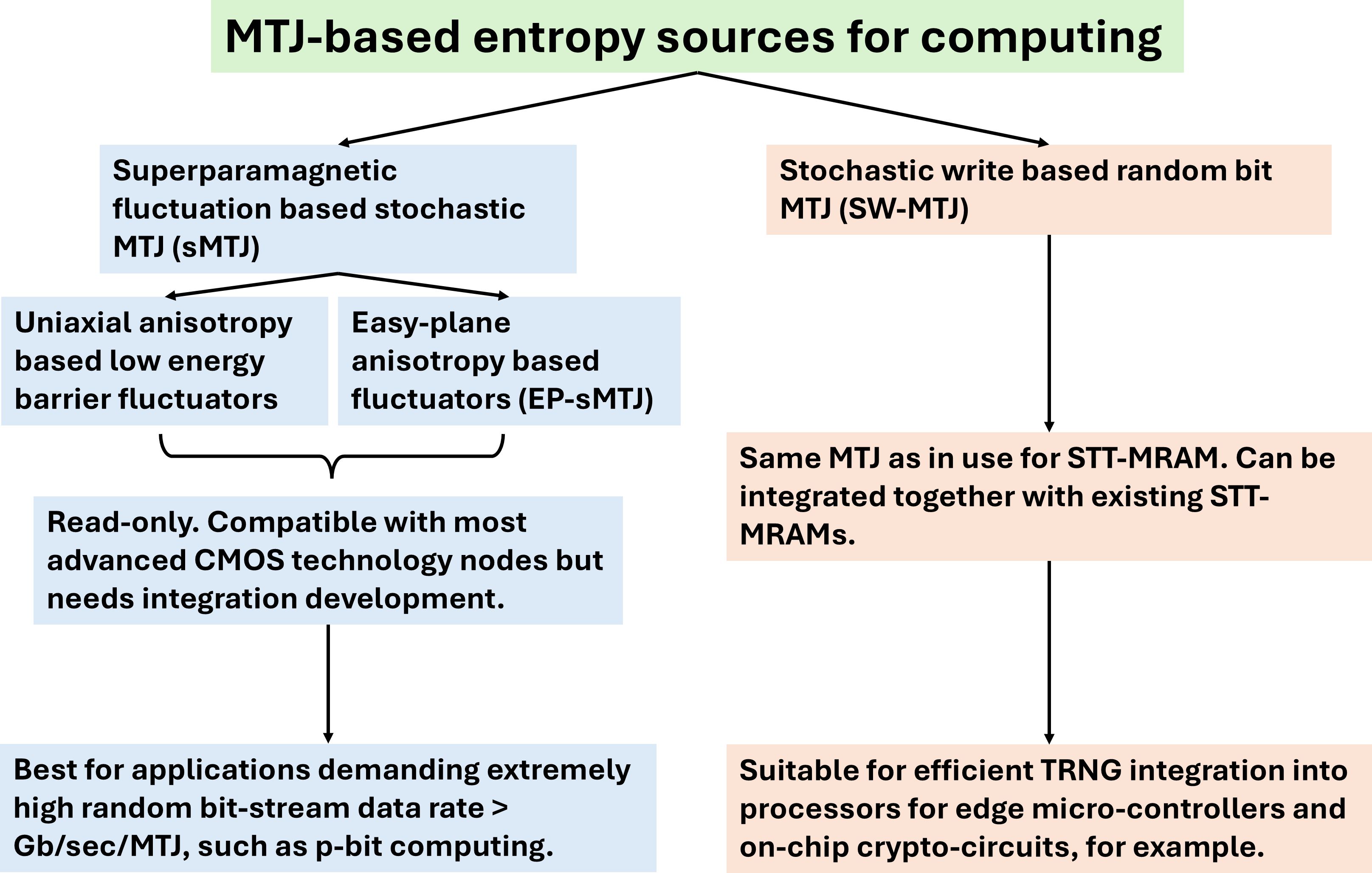}
\centering
\caption{Overview of MTJ-based entropy sources, comparing superparamagnetic fluctuation devices (sMTJs), their subtypes, and stochastic write devices (SW-MTJs), integration compatibility, and typical application domains.}
\label{figf0}\end{figure}
Figure~\ref{figf0} shows the classes of MTJ-based entropy sources. Two leading approaches are: (1) MTJs operating in the superparamagnetic regime (sMTJs), in which thermal agitation drives magnetic fluctuations of a MTJ magnetic free layer, and (2) Magnetically stable MTJs operated at reduced write voltage or pulse duration to intentionally produce probabilistic switching characteristics, referred to as a stochastic write MTJ (SW-MTJ).

For superparamagnetic MTJs, two general categories can be distinguished, as shown on the left-hand side of Fig.~\ref{figf0}. The first, referred to here as uniaxial sMTJs, employs a low-barrier uniaxial anisotropy MTJ in which the free-layer moment—and thus the tunnel magnetoresistance—fluctuates between two well-defined states: parallel and antiparallel, corresponding to low- and high-resistance levels. The second, referred to as easy-plane sMTJs, employs an MTJ with strong easy-plane magnetic anisotropy, confining thermal fluctuations of the magnetic moment essentially to the film plane. In this case, at any instant, ideally, the moment has equal probability of pointing in any in-plane direction, leading to a tunnel magnetoresistance as analog white-noise signal up to a  cut-off frequency $f_\mathrm{c} \sim \gamma H_\mathrm{p}$ associated with the easy-plane anisotropy field $H_p \sim 4\pi M_s$ where $M_s$ is the saturation magnetization of the free-layer (FL) of the MTJ\cite{2018136}.  This analog resistance-fluctuation signal can then be digitized or otherwise processed to generate a random bit stream. 

These two approaches --- the read-only, thermal-fluctuation based superparamagnetic MTJ (sMTJ) and the stochastic write MTJ (SW-MTJ) --- as illustrated in Fig.~\ref{figf0}, are suited for different applications and have different technology-integration attributes.

The sMTJ approach operates by passively reading the device resistance to obtain entropy, consuming minimal power. Single-device demonstrations have achieved bit rates of 0.5--1~Gb/s per sMTJ. Because it does not require a current-intensive write operation, the sMTJ is compatible with the most advanced CMOS nodes, provided BEOL MTJ fabrication can be optimized for consistent performance metrics (discussed below). This makes it particularly suitable for circuits and systems that demand the highest possible random bit rate at minimal power, such as large-scale probabilistic computing\cite{2025087,2024104,2024084,2023175}.

The SW-MTJ approach operates by actively strobing the MTJ with write pulses, making it inherently more drive-current and power intensive and somewhat slower ($\lesssim 10\times$) in maximum achievable bit rate than the sMTJ. Its main advantage is that it requires only minor circuit modifications relative to standard STT-MRAM technology, often using the same pMTJ devices (or slightly modified versions, {\em e.g.}, medium energy barrier free layers). The same pMTJ process can be shared between STT-MRAM and SW-MTJ-based probabilistic bit (p-bit) circuits, enabling both to coexist on the same chip. This makes the SW-MTJ well suited for applications such as integrated cryptographic functions that frequently require substantial volumes of random bits but do not demand the highest possible bit rate.

Below, we explore in greater depth the distinctive characteristics of each approach. We review the underlying device physics and current technological status, and identify key materials, devices, and engineering challenges that must be addressed to advance these MTJ-based entropy sources toward large-scale deployment.

\section{Superparamagnetic MTJ as entropy sources}\label{sMTJ}
\begin{figure}[tbp!]
\includegraphics[width=3.2in]{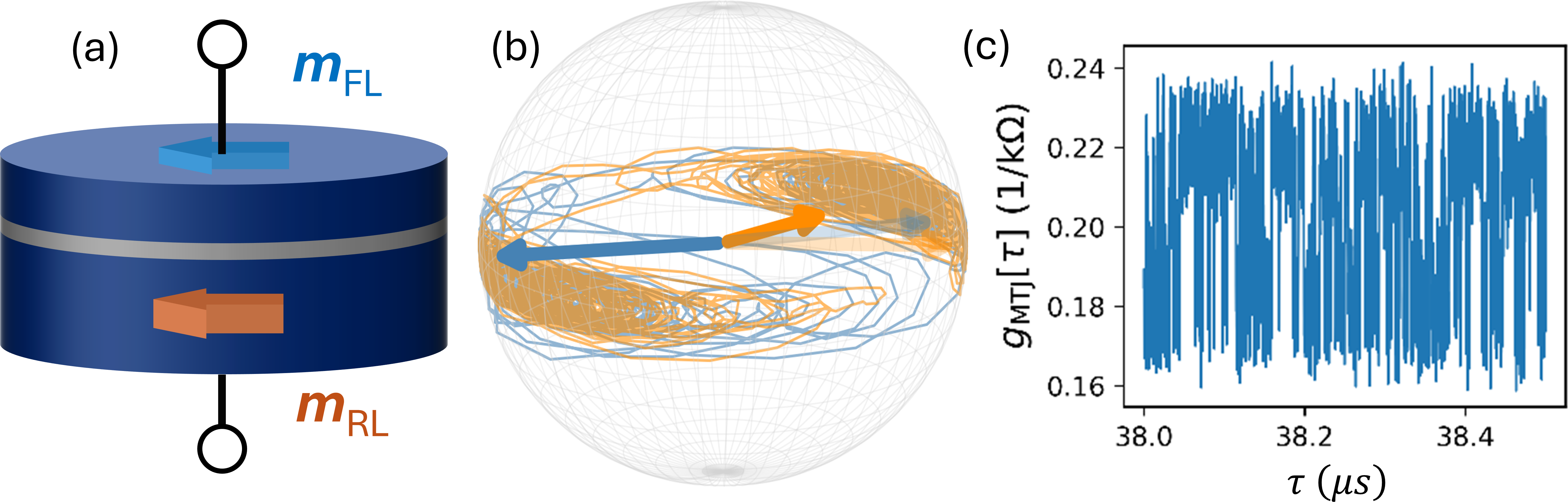}
\centering
\caption{Superparamagnetic MTJ (sMTJ) as an entropy source. 
(a) Schematic of a MTJ, in which the relative orientation of the free-layer magnetization $m_\mathrm{FL}$ and the reference-layer magnetization $m_\mathrm{RL}$ determines the electrical conductance. 
(b) Simulated thermal fluctuations of $m_\mathrm{FL}$ (in blue) and $m_\mathrm{RL}$ (in orange) on the unit sphere when the anisotropy energy barriers for $m_\mathrm{FL,RL}$ magnetization rotation are sufficiently low. 
(c) Example of time-dependent conductance fluctuations $g_\mathrm{MTJ}(\tau)$ arising from such thermal dynamics. 
The trajectories in (b) are obtained from finite-temperature Landau–Lifshitz–Gilbert simulations with a stochastic field---a Langevin field---for illustration purposes, whereas the conductance trace in (c) is measured from an actual MTJ device. Adapted from Ref.~\cite{2023044}.}
\label{figf1}\end{figure}

A superparamagnetic magnetic tunnel junction (sMTJ) is an MTJ in which the variation in magnetic anisotropy energy of the free-layer magnetization with orientation is comparable to or smaller than thermal energy $k_BT$. The magnetization can thus fluctuate over a wide range of angles, producing tunnel magnetoresistance variations spanning the full range between the parallel and antiparallel states, as conceptually illustrated in Fig.~\ref{figf1}. The specific form of the energy potential determines the entropy-generation characteristics of sMTJs.

\subsection{Low uniaxial anisotropy barrier based sMTJs}

A straightforward sMTJ design is one with a very low thermal-reversal anisotropy energy barrier, while retaining uniaxial anisotropy, usually perpendicular to the film plane. This design departs minimally from an MTJ optimized for memory applications: the net perpendicular anisotropy energy is reduced to a value comparable to the operating thermal energy. Thermal fluctuations then naturally drive telegraph-like switching of the magnetic moment between the two minima of the uniaxial anisotropy potential. Because this design requires only lowering the anisotropy energy barrier of a conventional STT-MRAM cell, such devices have been among the first sMTJ candidates explored experimentally for random-bit generation~\cite{2019164}.

There are, however, fundamental constraints to this approach arising from basic device and thermal physics. It is well established~\cite{2003004,2013104,2020088,2019122} that the mean escape lifetime, $\tau_\mathrm{esc}$---related but not identical to the autocorrelation time $\tau_\mathrm{ACF}$---of a low-barrier uniaxial superparamagnet scales asymptotically in the macrospin limit as
\begin{equation}
\tau_\mathrm{esc} \sim \frac{m}{\alpha \gamma k_B T},
\label{A2.1}
\end{equation}
where $m = M_s V_\mathrm{FL}$ is the magnetic moment of the free layer, with magnetization $M_s$, volume $V_\mathrm{FL} = (\pi/4)a^2 t$, diameter $a$, and thickness $t$. Here $\gamma \approx 2 \mu_B/\hbar$ is the gyromagnetic ratio, $\alpha$ is the Gilbert damping, $k_B$ the Boltzmann constant, and $T$ the operating temperature.

This asymptotic expression, given by Eq.~74 of Ref.~\cite{2013104}, represents only the mean lifetime for thermally activated hopping. The actual autocorrelation time of the resulting time sequence is slightly shorter according to finite-temperature Landau–Lifshitz–Gilbert simulations with a Langevin field. The addition of an orthogonal easy-plane anisotropy~\cite{2021047,2022003} or an orthogonal magnetic field along the hard axis~\cite{2023080,2022106} has also been considered to decrease the escape time, even down to ns level shown by some recent experiments\cite{2025115}.  

This linear dependence of $\tau_\mathrm{esc}$ on $m/k_B T$ implies that the temporal characteristics of the random bit stream will vary sensitively with temperature, an undesirable property for applications requiring stable operation across a wide temperature range.

The dependence expressed in Eq.~\ref{A2.1} also imposes a practical lower bound on the temporal spacing of random bits, $\tau_\mathrm{ACF}$, for CoFeB-type MTJs. In practice, $\tau_\mathrm{ACF}$ cannot be reduced below a fraction of a microsecond because the magnetic volume, set primarily by the free-layer thickness and diameter, can only be decreased to a limited extent.\footnote{A certain thickness, typically $>$1~nm, is required to achieve sufficiently large tunnel magnetoresistance (TMR), and an MTJ in-plane diameter of order $\sim 30$~nm or larger is needed for robust lithography processing in CMOS integration.} Several single-device-level experiments~\cite{2021047,2022003,2023080,2022106,2023152} have confirmed this behavior, with results generally consistent with the considerations discussed above.

In principle, materials with a lower $M_s$ than CoFeB could be explored for this purpose. Such materials would need to simultaneously deliver large TMR, a target junction resistance–area product (e.g., $r_A \sim 10~\Omega\,\mu\mathrm{m}^2$), and a very low anisotropy energy profile (see Sec.~\ref{status}.1). Realizing these properties would also require overhauling device characterization, integration, and validation procedures for circuit implementation—adding significant cost to commercial technology development alongside the expense of new materials research. Nevertheless, successful identification and integration of such materials might enable faster fluctuation dynamics and improved scalability for sMTJ-based entropy sources and expand their applicability in advanced computing.

\subsection{Easy-plane anisotropy based sMTJs for sub-nanosecond random bits}\label{EP-sMTJ}

Another sMTJ approach employs a free layer (FL) with a strong easy-plane anisotropy while remaining isotropic in-plane. This configuration, referred to as an easy-plane stochastic MTJ (EP-sMTJ), has been shown in low damping macrospins~\cite{Sun18,2019122,2020088} to exhibit a thermal fluctuation autocorrelation time $\tau_\mathrm{ACF}$ of approximately  
\begin{equation}
\tau_\mathrm{ACF} \approx 0.42 \left( \frac{h}{2\mu_B} \right) \sqrt{\frac{V_\mathrm{FL}}{4 \pi k_B T}},
\label{A3.1}
\end{equation}
when the FL is modeled as a very thin film ($t \ll a$, where $a$ is the diameter) with a full demagnetization-induced $4\pi M_s$ easy-plane field. Here, $V_\mathrm{FL} = (\pi/4) a^2 t$ is the volume of the fluctuating FL, $M_s$ is the saturation magnetization, and the system is assumed to be in the limit of large easy-plane anisotropy energy, such that $(2\pi M_s^2) V_\mathrm{FL} / k_B T \gg 1$, i.e. the moment's thermal fluctuation stays largely in-plane.

EP-sMTJs have been the subject of several experimental studies~\cite{2021034,2021047,2022061,2023044,2023171}, with demonstrations~\cite{2021034} showing fluctuation autocorrelation times as short as $\tau_\mathrm{ACF} < 5$~ns in bench-top measurements using 50~$\Omega$ coaxial transmission lines. These results likely do not represent the fundamental $\tau_\mathrm{ACF}$ limits imposed by material properties, but are instead constrained by the test methodologies and device implementations employed thus far. 

\begin{figure}[tbp!]
\includegraphics[width=2.8in]{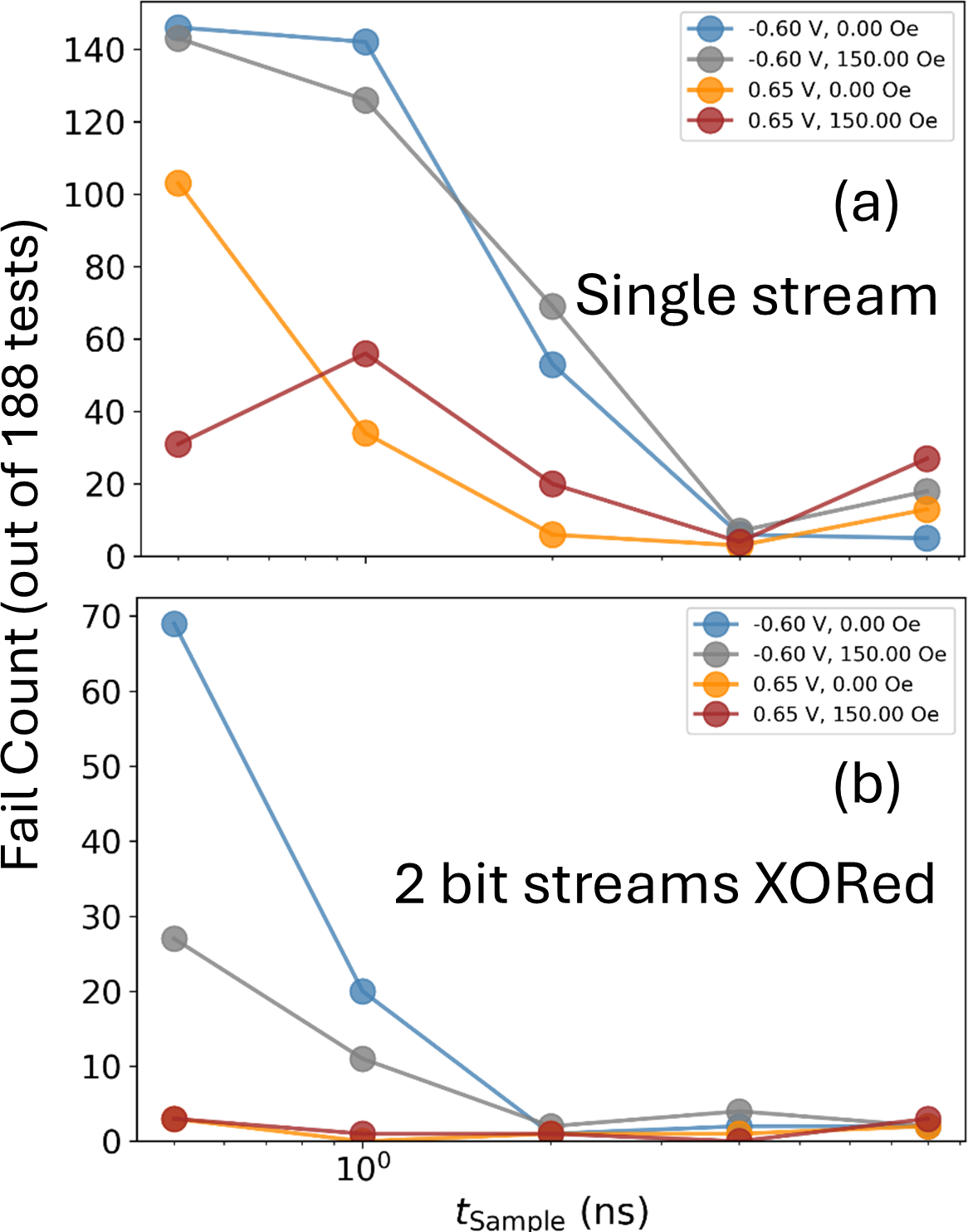}
\centering
\caption{Sampling-time dependence of fail counts from the NIST~SP800-22r1a test suite (out of 188 tests~\cite{NISTPub}) for an $\simeq~35$ nm diameter EP-sMTJ with a synthetic antiferromagnetic free layer, measured under different bias voltage and magnetic field conditions. (a) Single bit stream without post-processing. (b) Output obtained by applying a single XOR operation to two independently acquired bit streams. Data from~\cite{2023171}.
}

\label{figf1b}\end{figure}

Gps rate random bit stream has also been demonstrated using EP-sMTJs at the single-device level, providing an existence proof~\cite{2024129,2023171,2023044,2021034}. Figure~\ref{figf1b} shows an EP-sMTJ of approximately 35~nm in diameter with a synthetic antiferromagnetic free layer, producing $\sim$0.5~Gb/s bit streams that pass the NIST~SP800-22r1a (and NIST~SP800-90B, not shown) randomness tests~\cite{NISTPub,NISTSP80090B}. Combining two independently acquired bit streams with a single XOR operation yields $>$1~Gb/s bit streams that also pass these NIST tests. Devices incorporating antiferromagnetically coupled composite free layers thus appear to perform better in generating high-speed bit streams that meet NIST criteria. Furthermore, more symmetric EP-sMTJs—in which both the FL and RL undergo superparamagnetic fluctuations—tend to produce random states less sensitive to voltage-bias-induced spin torque, which can otherwise shift the mean value of the fluctuating magnetoresistance~\cite{2020127B,2024129,2023171}.

While the structure of an EP-sMTJ is compatible with large-scale CMOS back-end integration as in STT-MRAM technology, initial explorations have revealed technical challenges. Foremost among these is an unintended and uncontrolled residual uniaxial anisotropy within the film plane, observed when using processes, materials, and device designs derived from STT-MRAM fabrication. Although relatively small from a memory-operation perspective (on the order of several $k_B T$ at ambient temperature), this residual anisotropy varies in both magnitude and orientation from device to device on the same wafer and even within the same die. Such variability must be mitigated; otherwise, EP-sMTJs would exhibit excessive spread in their fluctuation characteristics, limiting their practicality for circuit applications. Addressing this challenge will likely require targeted materials optimization, refined process control, and device design strategies that suppress or compensate for in-plane anisotropy variations.

The leading suspect for the unintended and uncontrolled in-plane anisotropy is magnetostriction coupling to a residual strain field near or at the MTJ’s free-layer/tunnel-barrier region. Such strain is difficult to eliminate entirely due to multiple aspects of MTJ BEOL integration, including post-deposition annealing–induced crystallization required to achieve high TMR, and the use of material combinations that provide large TMR but also exhibit high magnetostriction coefficients—both in bulk CoFe and at its interface with the MgO tunnel barrier.

For example, amorphous (CoFe)$_{80}$B$_{20}$ in its bulk state has a magnetostriction coefficient of approximately $\lambda_s \approx 2\times10^{-5}$, while crystalline CoFe exhibits values at least three times higher~\cite{CFBMagnetostriction,2015169,2023002}. In ultrathin CoFe/MgO films, interfacial effects further enhance the magnetostriction~\cite{2015040,2023029}. MTJs incorporating such materials are fabricated in a BEOL environment where the residual film stress can approach the plastic deformation threshold, $\sigma \approx 0.1$--$1$~GPa~\cite{2024042}. Further, this stress may be released and reintroduced multiple times in an uncontrolled manner during various BEOL lithographic processing steps.

The anisotropy energy arising from magnetostriction and the stress-field strength can be estimated as~\cite{CFBMagnetostriction,2023044,2023171}  
\begin{equation}
E_b \approx \frac{3}{2} \lambda_s \sigma \left( \frac{\pi}{4} a^2 t \right),
\label{E1}
\end{equation}
where $E_b$ is the in-plane uniaxial anisotropy energy of an sMTJ free-layer disk of diameter $a$ and thickness $t$, and $\lambda_s$ is the magnetostriction coefficient. The stress-field magnitude is given by $\sigma \approx E_y \varepsilon$, where $\varepsilon \sim \delta l / l$ in the one-dimensional limit and $E_y$ is the Young’s modulus of the relevant material. For reference, $1$~GPa corresponds to $10^{10}$~erg/cm$^3$.

Using Eq.~\ref{E1} with representative parameters for a CoFeB-type MTJ free layer—$M_s \sim 600$~emu/cm$^3$, $\lambda_s \sim 2\times10^{-5}$, and $\sigma \sim 0.1$~GPa—for a junction with $a = 35$~nm and $t = 2$~nm, yields $E_b \sim 1.4\,k_B T$ and $H_k \sim 100$~Oe. These values are consistent with the order of magnitude of the unintended in-plane anisotropy observed in otherwise isotropic EP-sMTJs~\cite{2023044,2023171}.

Reducing uncontrolled in-plane magnetostriction is currently a major technical challenge for developing viable materials in sMTJ design. The effective $\lambda_s$ of the sMTJ must be lowered by approximately an order of magnitude, and/or the typical uncontrolled back-end-of-line stress field reduced by about a factor of five, while maintaining a tunnel magnetoresistance as high as possible (and certainly $\gg 100\%$). Achieving these targets will require substantial materials and process co-optimization before realistic BEOL integration of sMTJs with CMOS technologies can be realized.

In addition, all full-bandwidth, sub-nanosecond EP-sMTJs demonstrated to date, such as those shown in Fig.~\ref{figf1b}, have been conducted with direct interfaces to 50~$\Omega$ high-speed test circuits~\cite{2021034,2021047,2023044,2023171}. Such low-impedance loading can introduce spin-torque effects that alter the thermal fluctuation statistics of the EP-sMTJ, potentially causing measurement-derived behavior to differ from that under the higher-impedance conditions required for CMOS-transistor integration. Comprehensive characterization of CMOS-ready EP-sMTJs at high impedance, with minimal current loading, will require on-chip active test circuits.

\section{Stochastic write MTJs as random bit generators}\label{strobe-write}

\begin{figure}[tbp!]
\includegraphics[width=3.2in]{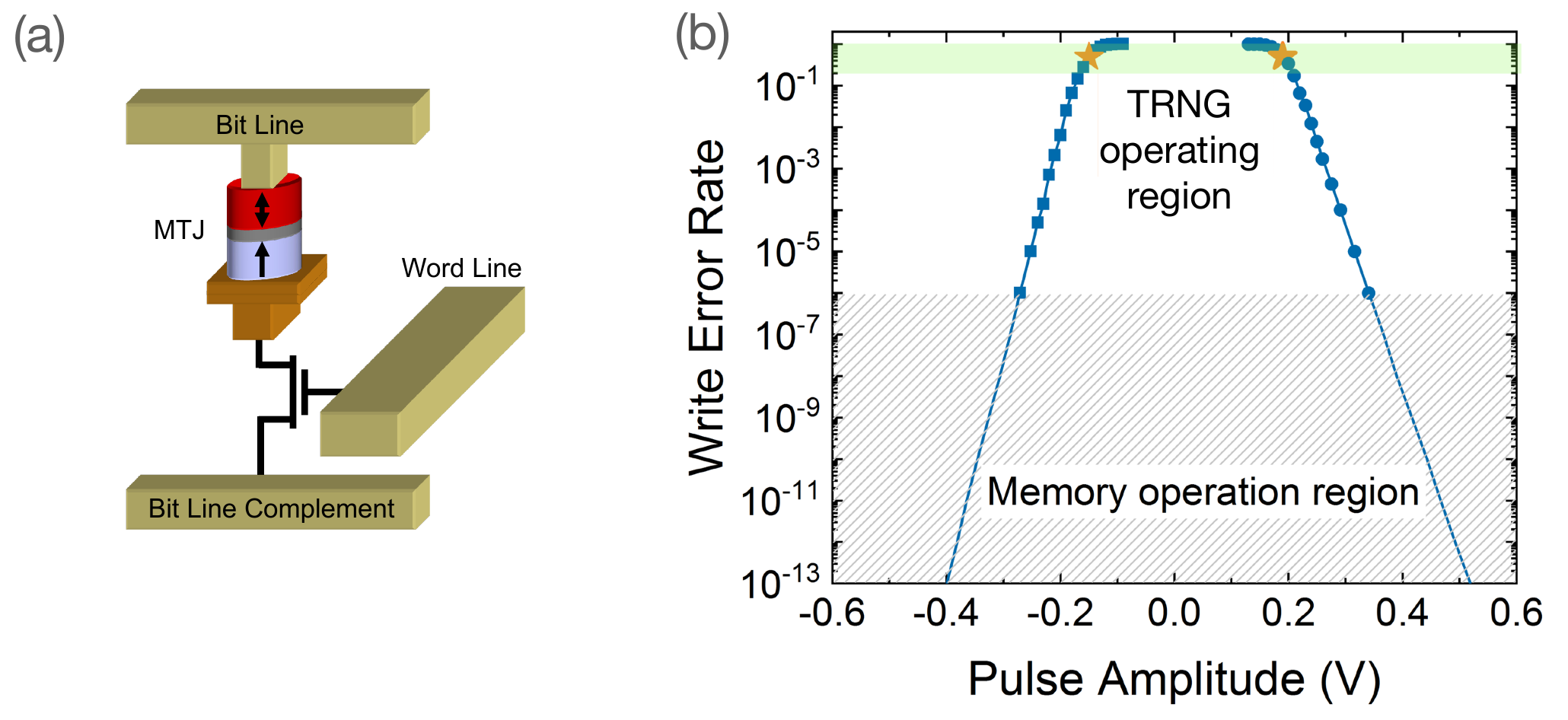}
\centering
\caption{Stochastic-write MTJ (SW-MTJ) true random-bit generator. (a) Schematic of a 1-transistor/1-MTJ memory cell, where the select transistor enables voltage-pulse application across the MTJ for write operations. Adapted from Ref.~\cite{2025069}. (b) Write-error rate (WER) characteristics as a function of pulse amplitude at fixed pulse duration, highlighting in green the operating region for random-bit generation in contrast to the low-write-error regime (the hatched area) required for memory operation. Adapted from Ref.~\cite{2024047}.}
\label{figf2}\end{figure}

Another approach to using STT-MRAM–compatible MTJs for random-number generation leverages the probabilistic nature of STT switching at finite temperature. Fig.\ref{figf2} gives an illustration. In the STT-MRAM context, this is commonly referred to as the “write-error-rate” (WER) characteristic of a given MTJ.  In conventional STT-MRAM operation, the write voltage is set sufficiently high to achieve STT switching with an extremely low error rate~\cite{2011121,2011136A,2014031,2016085,2016192,2016084,2023046}. 
For random-bit generation, by contrast, the MTJ is driven with reduced voltage and/or shorter pulse duration, so that each write attempt succeeds only with finite probability. This yields random bit states that can be read using standard MTJ–CMOS circuitry, but with operating points chosen for rapid write/read cycling rather than the high write reliability and read-out accuracy required in memory applications.

The WER characteristics of MTJs at the device level have been extensively studied and are quantitatively understood~\cite{2011121,2014011,2016085,2021191,2022120,2023123,2024056,2024058}. Remaining challenges include fully assessing the stability and statistical variance of operating points—both in repeated single-device operation, where long-term and temperature-dependent drift can occur, and across device-to-device variability. Recent experiments demonstrate that operating stochastic-write MTJs with nanosecond-duration pulses, in the ballistic limit, greatly reduces sensitivity to temperature variations~\cite{2025114}, while modeling indicates reduced susceptibility to process-voltage-temperature variations and device-to-device spread~\cite{2024056}. In this regime, in the macrospin approximation, the temperature sensitivity of the probability bias close to $p=0.5$ is given analytically by
\begin{equation}
\frac{dp}{dT} = \frac{\ln 2}{2T},
\label{Eq:dpdT}
\end{equation}
which depends only on the absolute temperature and, remarkabkly, not on device-specific material parameters. At room temperature ($T=300$ K), this expression gives $dp/dT = 0.0016$ K$^{-1}$, representing the lower limit of temperature sensitivity in the ballistic switching limit.

SW-MTJs have been shown to generate random bits at rates exceeding 0.1 Gbps, with long-term probability bias drift correlated with changes in MTJ electrical resistance~\cite{2025103}. These findings highlight that maintaining stable magnetic and electrical characteristics is essential for reliable long-term performance. At the same time, circuit- and system-level optimization remains necessary to determine the ultimate power-speed-performance trade-offs of SW-MTJ–based TRNGs.

\section{Performance metrics for entropy sources}
\subsection{Speed, power and cell-size efficiency}

Physical-entropy-based true random number generators (TRNGs) are generally many orders of magnitude more efficient in both size and energy consumption than standard processor-based solutions. As a baseline for estimating the cost of random bits in practical computational workloads, we refer to a recent survey~\cite{2024032}. This study reported that algorithm-driven pseudorandom number generation (PRNG) at data rates of order 1 Gbps per stream achieves energy efficiencies of approximately 30 nJ/bit on CPUs and about 3 nJ/bit on GPUs.

Table~\ref{table1} provides a comparative overview of several electronic random-bit generation methods based on physical entropy sources and developed for computational use. These include entropy derived from dynamic CMOS circuit chaos~\cite{2021123}, thermal-noise-based CMOS circuits~\cite{2021132}, integrated tie-down diode noise~\cite{2023177,2024121}, and single-photon avalanche diodes (SPADs)~\cite{2023174,2024122,2023173}. While effective, these approaches all involve trade-offs in speed, integration complexity, or energy efficiency. Not included in this comparison are opportunistic entropy-harvesting methods that, although limited to relatively low data rates, remain adequate for cryptographic applications—for example, exploiting the dark noise of CMOS image sensors commonly found in mobile devices~\cite{2024120}.

\begin{table*}[tbp!]
\centering
\caption{Comparison of the performance of several random number generator types based on physical entropy sources. Abbreviations: CCMOS—chaotic CMOS (130\,nm technology); TCMOS—thermal-noise CMOS (65\,nm node); ITD—integrated tie-down diode; SPAD—single-photon avalanche diode; DC-OC—discrete-component optical cavity.}
\label{table1}
\begin{ruledtabular}
\begin{tabular}{lcccccccc}
& {\bf EP-sMTJ}\footnote{Estimated in this work assuming $R_\mathrm{MTJ}=10$ k$\Omega$, $V_\mathrm{dd}=$0.8~V, $I_\mathrm{read}=$10$\mu$~A, $I_\mathrm{strobe}=$50$~\mu$A, $\tau_\mathrm{read}=2$~ns, $\tau_\mathrm{strobe}=1$~ns, excluding RC and circuit overheads. Size entries reflect only the entropy-sourcing device itself and do not include the supporting circuit footprint, which require one or more front-end CMOS transistors to address and operate the MTJs.} 

& {\bf SW-MTJ}$^\text{a}$ &{\bf CCMOS}\cite{2021123} & {\bf TCMOS}\cite{2021132} & {\bf ITD}\cite{2024121} & {\bf SPAD}\footnote{Cell estimates do not include illumination control which is necessary for SPAD functionality and tuning.}\cite{2024122} & {\bf DC-OC}\footnote{QUSIDE Ruby$^\text{TM}$ N1 discreet-chip based. Product spec: https://quside.com/product/ruby-series-qrng-chipset/} \\
\hline\\
Cell current ($\mu$A) & $\sim$ 20 & $\sim$ 50 & N/A & N/A & 51.7 & N/A & N/A \\
Energy efficiency (pJ/b) & 0.03 & 0.08 & 4.37 & 0.36 & 0.28 & N/A &170 \\
Speed (Gbps/cell) & $\gtrsim $1 & $\lesssim 0.1 \sim 0.5$  & 0.4 & 0.1 & 0.15 & $\sim 1$ &$\lesssim$ 1\\
Cell size ($\mu$m$^2$) &$ \sim 0.1\times0.1$  &$ \sim 0.1\times0.1$&$ 400\times400$ & $ 40\times50$ & 203 & $37\times35$ &  $5\times5$ mm$^2$\\
\end{tabular}\end{ruledtabular}\end{table*}

Table~\ref{table1} illustrates that true random number generators (TRNGs) based on physical entropy sources are generally orders of magnitude more efficient in both size and energy consumption than processor-based solutions. Among them, entropy sources such as EP-sMTJs and SW-MTJs provide best-in-class performance, delivering high bit-rate random bit streams (100 Mbps to Gbps) with minimal circuit overhead and the ability to integrate directly alongside logic-processing circuits. These characteristics make physical-entropy TRNGs, including sMTJs and SW-MTJs, particularly attractive for random-number-intensive workloads, provided that the challenges of hardware cost and system-level integration can be addressed.

In evaluating bit stream quality for cryptographic applications, established standards provide rigorous benchmarks. A common starting point is to apply standard test suites, such as NIST SP 800-22r1a and NIST SP 800-90B~\cite{NISTPub,NISTSP80090B}, to unbiased stochastic bit streams  ($p = 0.5$). For other applications, including Monte Carlo modeling and emerging paradigms such as p-bit computing, the most relevant performance metrics remain less well defined. Beyond control of the mean, important factors such as autocorrelation length, energy efficiency, bit rate, and device scalability have yet to be standardized.

\subsection{Hardware control of bit stream statistics}
For applications such as probabilistic-bit computing, hardware needs to be designed and integrated to control the statistical properties of the random bit stream, including its mean and distribution. Several approaches are being investigated to provide this functionality.

The simplest approach combines analog circuit design with a fixed entropy source, such as an EP-sMTJ, to control the mean of the output bit stream~\cite{2018146,2018149,2019164}. This method places minimal requirements on the EP-sMTJ, as each bit stream-generating cell can employ an identical but independent entropy source. Nonlinear circuit elements define the digitization threshold, thereby providing control of the bit stream mean.

A potentially more compact solution in terms of circuit footprint is to employ device-physics-based designs that incorporate a third terminal on the entropy-generating component to control its fluctuation mean. Spin-orbit torque–based three-terminal devices, for example, have been proposed as promising candidates~\cite{2017086,2017094}.

A meaningful system-level performance comparison will ultimately require scaling studies that incorporate these design specifics---an important task that remains to be done.

An important component- or circuit-level metric is the timescale over which the random bit stream’s distribution evolves. This response time sets the effective operating speed of asynchronous p-bit networks~\cite{2020089} and motivates the need for high bit-rate p-bit streams in such applications. In circuit-controlled bit streams, the relevant timescale is primarily governed by the response of the analog threshold to the control signal, which can be evaluated with established circuit models. In contrast, when distribution control is implemented directly at the device level~\cite{2017086,2017094}, additional device-physics factors come into play. For example, in spin-torque control of a superparamagnetic fluctuator’s distribution mean, materials properties and device operating conditions determine the intrinsic timescales~\cite{2022130}. Such device-based approaches also place greater demands on reproducibility and stability of fluctuation behavior and controllability, increasing fabrication and integration challenges. Nevertheless, some of these variations may be mitigated through application-specific, hardware-aware learning routines~\cite{2022061}. In sum, circuit-level approaches offer simplicity and well-understood design frameworks, whereas device-level approaches promise compactness and direct control but face greater challenges in reproducibility and manufacturability.

For SW-MTJs, the bit stream probability can be tuned by adjusting the amplitude or duration of the strobe-write pulse. The underlying device physics and bit stream statistics are well established and have been experimentally demonstrated~\cite{2022120,2023123,2024010,2024056,2025114,2025103,2024058}. In this case, the control response time is determined primarily by the full write–read cycle required for random-bit generation.

\section{Compatibility with CMOS}\label{status}
The use of sMTJs and SW-MTJs for random-number-generation hardware is attractive for a variety of applications, with their distinct characteristics influencing the statistical quality of the resulting bit streams. These differences were briefly outlined earlier in connection with Fig.~\ref{figf0}. In this section, we expand on that discussion to provide a broader perspective on their compatibility with CMOS technologies.

\subsection{\label{sMTJstatus}Read-only operation enables sMTJ at advanced CMOS nodes}

The passively read easy-plane sMTJ (EP-sMTJ) is by far the fastest and most efficient approach to random bit stream generation in terms of energy and cell size. It is also inherently scalable, remaining fully compatible with the most advanced CMOS technology nodes with no constains in terms of transistor scaling.

This advantage arises from the EP-sMTJ’s `read-only' operation. The device produces essentially white noise with bandwidths extending beyond the gigahertz range and resistance fluctuations exceeding a factor of two. It further offers exceptional impedance design flexibility,\footnote{Because the tunneling conductance depends exponentially on barrier thickness, a wide range of impedances can be engineered to meet circuit requirements.} allowing perfect impedance matching to any generation of CMOS front-end technology. This unique combination of wide bandwidth, large output amplitude, and tunable impedance makes the EP-sMTJ particularly well suited as a scalable entropy source.

Unlike STT-MRAM or even SW-MTJs, the EP-sMTJ approach does not require any CMOS-driven switching. In conventional designs, the CMOS-controlled ‘write’ operation sets limits on circuit write currents and constrains deployment at the most advanced CMOS nodes, where available on-state current decreases with shrinking transistor dimensions. By eliminating the need for a write operation, sMTJs remain fully compatible with advanced CMOS technologies and are well suited for back-end integration.

One of the main challenges for EP-sMTJ devices is mitigating unintended in-plane anisotropy that can arise in BEOL integration, as discussed earlier in Sec.~\ref{sMTJ}. For EP-sMTJs to achieve large-scale CMOS integration and system-level demonstrations, these anisotropies must be reduced by at least an order of magnitude—an outstanding task in materials development, device design, and process optimization.

\subsection{SW-MTJ is compatible with STT-MRAM}
The SW-MTJ, as discussed in Sec.~\ref{strobe-write}, has a distinct set of advantages and disadvantages compared to the EP-sMTJ. It requires a write operation, though at reduced drive voltage and/or shorter pulse duration. As a result, the cell-level peak write current remains constrained—below that required for reliable STT-MRAM operation but still likely exceeding 20$\mu$A. Moreover, higher target bit rates demand proportionally larger peak currents. This imposes scaling constraints on both the maximum achievable random bit rate and the CMOS nodes with which SW-MTJs can be integrated, as they are limited by the available transistor on-state current and the supply voltage $V_\mathrm{dd}$.\footnote{For discussion of projected transistor on-state current and $V_\mathrm{dd}$, see Refs.~\cite{2022080,2022038,2023046,2024105}.}

The key advantage of the SW-MTJ is that it can employ the same BEOL-integrated MTJ structures already used in STT-MRAM. This enables SW-MTJs to be directly incorporated into STT-MRAM-enabled logic chips, coexisting with memory arrays and fabricated within the same process flow. The only significant modification required is a redesign of part of the MTJ read/write circuitry to support random bit generation, giving SW-MTJs a unique practical advantage for integration compared to other TRNG approaches.

These attributes make SW-MTJs strong candidates for advanced on-chip cryptography in systems that already incorporate STT-MRAM, with minimal additional cost. This combination is particularly well-suited to current STT-MRAM technologies, such as edge-enabled low-power microcontrollers with persistent memory.

Challenges for SW-MTJs going forward include improving materials and device control to minimize variations—both long-term temporal drift and device-to-device variability---for reliable and robust integration. Recent studies have begun to address these issues at the single-device level~\cite{2024056,2024058,2025114,2025103}, providing baseline metrics for existing MTJ devices and identifying areas in need of optimization. Continued progress along these lines will be essential to advance SW-MTJs toward large-scale, reliable technology integration.

\section{Summary and outlook}
Nanostructured MTJs offer compact, low-power, and high-speed integrated entropy sources with strong potential to advance future computing architectures. As random-bit generators, MTJ-based entropy sources can reduce circuit cell size and improve power efficiency by orders of magnitude compared to current technologies, while delivering best-in-class data rates of approximately 1 Gbps per cell. These capabilities position MTJs as a promising foundation for next-generation secure, energy-efficient, and large-scale computing systems.

Two types of MTJ-based entropy sources have been demonstrated to date: the passive-read sMTJ approach and the strobe-write SW-MTJ approach. sMTJs impose no scaling constraints on CMOS technology and are therefore suitable for integration with the most advanced nodes. In contrast, SW-MTJs are directly compatible with today’s STT-MRAM technology and can be integrated into STT-MRAM–based microcontrollers with minimal design changes, requiring only the repurposing of a portion of the STT-MRAM die area for SW-MTJ operation.

At present, single-device sMTJs have demonstrated random bit rates of approximately 0.5 Gbps per MTJ, with bit streams successfully passing the NIST SP800-22r1a test suite~\cite{2023171}, albeit with considerable device-to-device variation. Similarly, strobe-write SW-MTJs have achieved bit rates of just greater than 0.1 Gbps, with bit streams passing the NIST SP800-22r1a tests~\cite{2025103} after a single XOR operation on two independent streams. Looking ahead, continued progress in reducing variability for sMTJs and minimizing post-processing for SW-MTJs will be pivotal milestones—advances that appear well within reach given current developments.

To fully realize their technological potential, passive-read sMTJs will require a controlled reduction of in-plane magnetic anisotropy by roughly an order of magnitude, while strobe-write SW-MTJs will demand improved control over both temporal stability and device-to-device variability.

\begin{figure}[tbp!]
\includegraphics[width=3.2in]{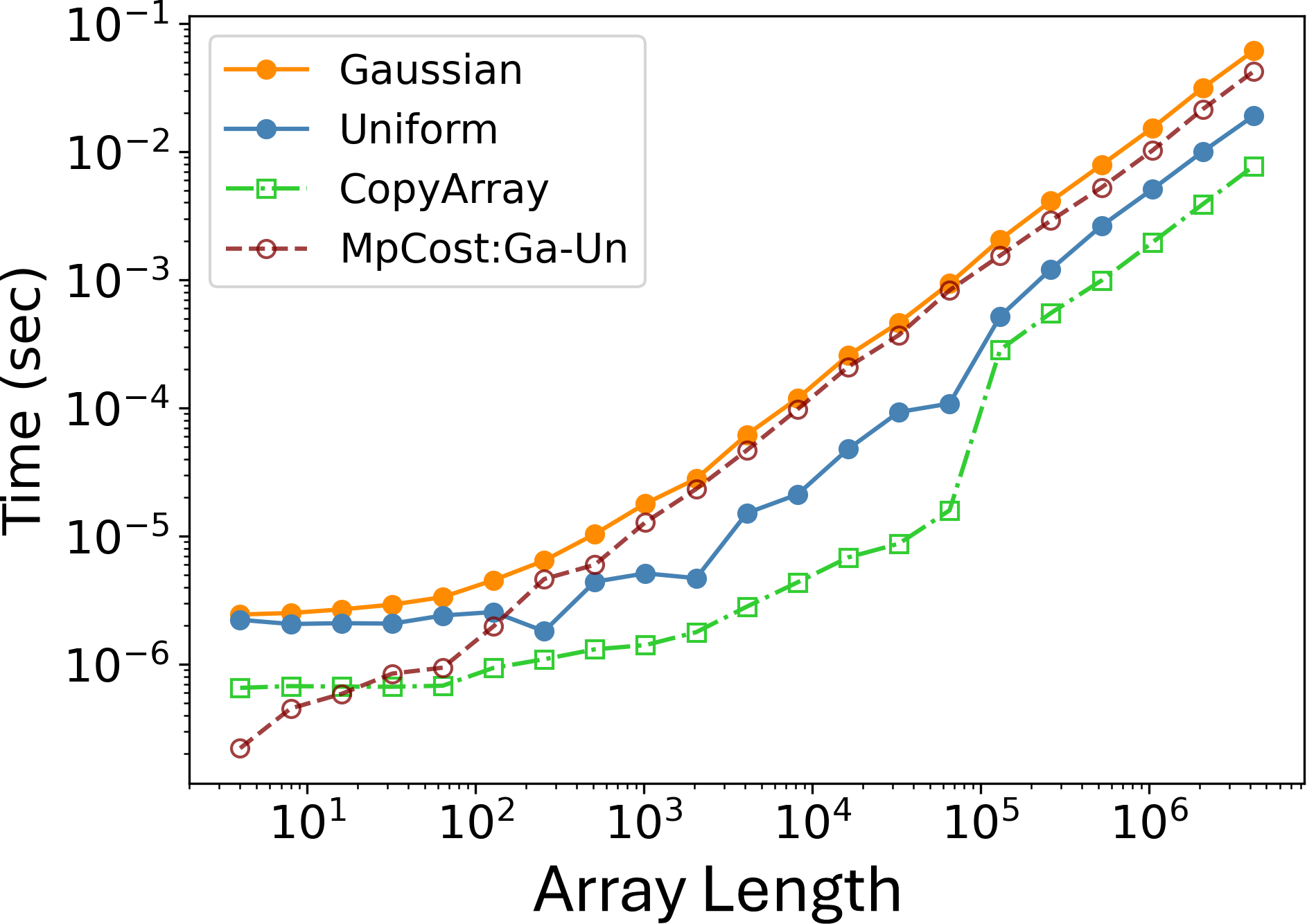}
\centering
\caption{Execution time for generating random arrays of 64-bit floating-point numbers as a function of array length, measured on a PC with an Intel 12th Gen i7-12700K CPU using NumPy 2.3.1~\cite{np231}. Results are shown for arrays with Gaussian-distributed values (``Gaussian'') and uniformly distributed values between 0.0 and 1.0 (``Uniform''). For comparison, ``CopyArray'' represents the time to copy a predefined array from global memory. ``MpCost'' indicates the additional computational cost of mapping a uniform sequence to a Gaussian distribution, estimated as the time difference between Gaussian and Uniform generation. This illustrates that the digital mapping step remains computationally intensive and dominates the cost of generating non-uniform random numbers.}
\label{figf3}\end{figure}

In addition to component- and cell-level performance, achieving a significant impact on system-level efficiency in computational tasks that rely heavily on random number generation will require effective methods for sampling from and controlling the statistical distribution of the generated numbers (see, for example, Ref.~\cite{2025104}).

Figure~\ref{figf3} provides one example of high-level system performance benchmarking. The data were generated on a PC with an Intel 12th generation i7-12700K CPU (1, 12, and 25 MB of L1, L2, and L3 cache, base clock 3.6 GHz) using NumPy 2.3.1. The figure compares execution times for generating pseudo-random uniform floating-point arrays versus arrays with Gaussian distributions. Producing a non-uniform random distribution, such as a Gaussian distribution, typically involves a two-step ``shaping'' process: first generating a uniformly distributed PRNG sequence---in this case using the PCG64DXSM bit generator~\cite{npPCG}—and then mapping it to the target distribution, here via the 256-step Ziggurat method~\cite{npZigguart}. Even with decades of optimization, this digital mapping step remains computationally intensive and dominates the cost of generating non-uniform random numbers, as seen in Fig.~\ref{figf3}. Thus, achieving truly efficient non-uniform random number sources for probabilistic computing requires more than reducing the cost of uniform random bit generation. It also calls for new approaches to distribution control that move beyond conventional digital implementations and demonstrate clear advantages in speed and energy efficiency under realistic performance benchmarks—a challenge that represents a key frontier for future TRNG research.
\subsection*{Acknowledgements}
The work at IBM was supported in part through a joint-development program with Samsung Electronics. 
The work at NYU was supported by the Office of Naval Research (ONR) under Award No. N00014-23-1- 2771 and in part by the National Science Foundation (NSF) under Award No. DMR-2105114. We acknowledge the contributions of Ahmed Sidi El Valli, Laura Rehm, Michael Tsao, Dairong Chen, Audre Dubovski, Troy Criss, and Andy Haas to the research at NYU on SW-MTJs.
\newpage 

\end{document}